\documentclass[final,preprintnumbers]{appolb}
\usepackage{epsfig}
\usepackage{axodraw}
\newcommand{\qq}{Q'} 
\newcommand{\re}{\mathrm{Re}\,}

\newcommand{\gev}{\,{\rm GeV}}
\newcommand{\tev}{\,{\rm TeV}}
\def\pb{\,{\rm pb}}
\begin{document}
\title{Timelike Compton Scattering at LHC
\thanks{Presented by J.Wagner at the ,,Excited QCD'' Conference, 8-14 February 2009, Zakopane}%
}
\author{B. Pire
\address{CPHT, {\'E}cole Polytechnique, CNRS, 91128 Palaiseau, France}
\and
L. Szymanowski, J. Wagner 
\address{Soltan Institute for Nuclear Studies, Ho\.{z}a 69, 00-681
Warsaw, Poland}
}
\maketitle
\begin{abstract}
Exclusive photoproduction of dileptons, $\gamma N\to
\ell^+\!\ell^- \,N$, is and will be measured in ultraperipheral
collisions at hadron colliders.
We demonstrate that the timelike deeply virtual Compton scattering (TCS)
mechanism  $\gamma q \to \ell^+\!\ell^- q $, where
the lepton pair comes from the subprocess      $\gamma q \to \gamma^* q $,
dominates in some accessible kinematical regions, thus opening a new 
way  to study generalized parton distributions (GPD)    in     the nucleon at
small skewedness. 

\end{abstract}
  
\section{Introduction}
General Parton Distributions (GPDs) \cite{Muller:1994fv} are subject to intense theoretical study and experimental effort. The best known process in which GPDs factorize from perturbatively calculable coefficient functions is Deeply Virtual Compton Scattering (DVCS). Even though the DVCS, being an exclusive process, is very different from Deep Inelastic Scattering (DIS) (see figure \ref{DVCS_DIS_1}) the description of the former can be easily understood as the generalization of description of the latter. Due to the optical theorem the cross section of DIS is proportional to the imaginary part of the Feynman diagram shown on the figure \ref{DVCS_DIS_2}a, which is a special case of the diagram describing the amplitude of DVCS (figure \ref{DVCS_DIS_2}b ).

GPDs are functions of three kinematical variables: longitudal momentum fraction $x$, skewedness $\xi$ and overall momentum transfer $t$. In the forward limit: $t, \xi \to 0$, GPDs reduce to PDFs. When integrated over $x$, GPDs reduce to elastic form factors. First moment of GPDs enter the Ji's sum rule for the angular momentum carried by partons in the nucleon. Fourier transform of GPDs to impact parameter space can be interpreted as ,,tomographic'' 3D pictures of nucleon, describing parton distribution in the transverse plane, for a given value of $x$.

In our work \cite{Pire:2008ea} we study the "inverse" process, $ \gamma p\to \gamma^* p$ at small $t$ and large timelike virtuality $\qq^{2}$ of the final state photon, timelike Compton scattering (TCS) \cite{TCS} which shares many features of DVCS. The possibility to use high energy hadron colliders as powerful sources of quasi real photons \cite{UPC} leads to the hope of determining sea-quark and gluon GPDs in the small skewedness region, which is an essential program complementary to the determination of the quark GPDs at lower energy electron accelerators. Moreover, the crossing from a spacelike to a timelike probe is an important test of the understanding of QCD corrections, as shown by the history of the understanding of the Drell-Yan reaction in terms of QCD.

\begin{figure}[t]
\begin{center}
\begin{picture}(200,60)(0,0)
\Text(0,55)[]{$e$}
\Text(80,55)[]{$e$}
\Text(25,30)[]{$\gamma^*$}
\Text(0,5)[]{$p$}
\Text(80,15)[]{$X$}
\ArrowLine(10,55)(40,45)
\ArrowLine(40,45)(70,55)
\ArrowLine(10,5)(40,15)
\Line(40,15)(70,8)
\Line(40,15)(70,15)
\Line(40,15)(70,22)
\Vertex(40,15){7}
\Photon(40,45)(40,15){2}{6}

\Text(100,55)[]{$e$}
\Text(180,55)[]{$e$}
\Text(125,30)[]{$\gamma^*$}
\Text(100,5)[]{$p$}
\Text(180,5)[]{$p$}
\Text(180,25)[]{$\gamma$}
\ArrowLine(110,55)(140,45)
\ArrowLine(140,45)(170,55)
\ArrowLine(110,5)(140,15)
\ArrowLine(140,15)(170,5)
\Vertex(140,15){7}
\Photon(140,45)(140,15){2}{6}
\Photon(140,15)(170,25){2}{6}
\end{picture}
\end{center}
\caption{Deep Inelastic Scattering (left) versus Deeply Virtual Compton Scattering (right).}
\label{DVCS_DIS_1}
\end{figure}
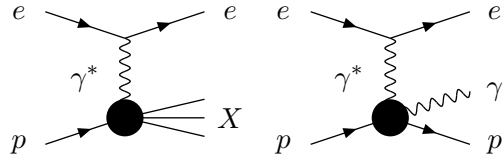
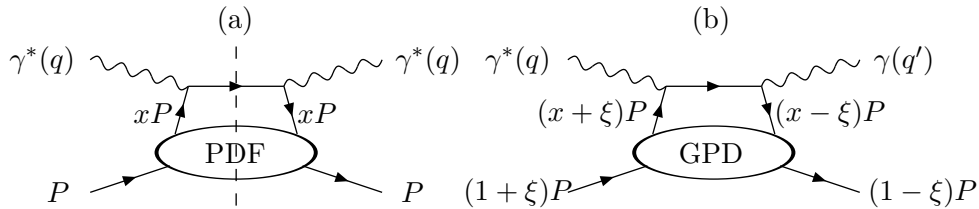
\begin{figure}[h]
\begin{center}
\begin{picture}(330,80)(0,0)
\Text(75,75)[]{(a)}
\Text(255,75)[]{(b)}
\DashLine(75,5)(75,65){5}
\Text(75,25)[]{PDF}
\Oval(75,25)(10,30)(0)
\ArrowLine(20,10)(50,20)
\ArrowLine(100,20)(130,10)
\ArrowLine(52,32)(57,50)
\ArrowLine(57,50)(93,50)
\ArrowLine(93,50)(98,32)
\Photon(20,60)(57,50){2}{4}
\Photon(93,50)(130,60){2}{4}
\Text(8,10)[]{$P$}
\Text(142,10)[]{$P$}
\Text(44,40)[]{$xP$}
\Text(106,40)[]{$xP$}
\Text(2,60)[]{$\gamma^*(q)$}
\Text(148,60)[]{$\gamma^*(q)$}
\Text(255,25)[]{GPD}
\Oval(255,25)(10,30)(0)
\ArrowLine(200,10)(230,20)
\ArrowLine(280,20)(310,10)
\ArrowLine(232,32)(237,50)
\ArrowLine(237,50)(273,50)
\ArrowLine(273,50)(278,32)
\Photon(200,60)(237,50){2}{4}
\Photon(273,50)(310,60){2}{4}
\Text(182,10)[]{$(1+\xi)P$}
\Text(335,10)[]{$(1-\xi)P$}
\Text(210,40)[]{$(x+\xi)P$}
\Text(300,40)[]{$(x-\xi)P$}
\Text(182,60)[]{$\gamma^*(q)$}
\Text(328,60)[]{$\gamma(q')$}
\end{picture}
\end{center}
\caption{Deep Inelastic Scattering cross section is given by the imaginary part of diagram (a). Amplitude of Deeply Virtual Compton Scattering is given by diagram (b).}
\label{DVCS_DIS_2}
\end{figure}

\section{Photoproduction of a lepton pair}
\begin{figure}[t]
\begin{center}
     \epsfxsize=0.35\textwidth
     \epsffile{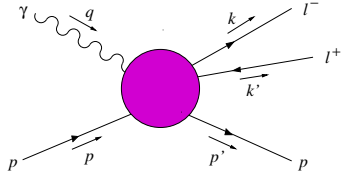}
\caption{\label{refig}Real photon-proton scattering into a lepton pair and a proton.}
\end{center}
\end{figure}

The physical process where to observe TCS, is photoproduction of a heavy lepton pair, $\gamma N \to \mu^+\!\mu^-\, N$ or $\gamma N \to e^+\!e^-\, N$, shown in Fig.~\ref{refig}. As in the case of DVCS, the Bethe-Heitler (BH)
mechanism 
contributes at the amplitude level. 
This process has a very peculiar angular dependence and overdominates the TCS process if
one blindly integrates over the final phase space. One may however choose kinematics where 
the amplitudes of the two processes are of the same order of magnitude, and either subtract the 
well-known Bethe-Heitler process or use specific observables sensitive to the interference of the two amplitudes. 
\begin{figure}[b]
\begin{center}
     \epsfxsize=0.55\textwidth
      \epsffile{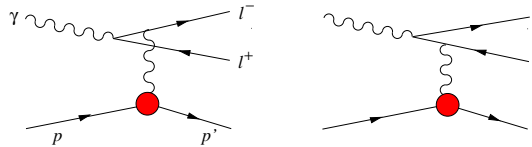}
\caption{The Feynman diagrams for the Bethe-Heitler amplitude.}
\label{bhfig}
\end{center}
\end{figure}
The Bethe-Heitler amplitude is  calculated from the two Feynman
diagrams in Fig.~\ref{bhfig}. 
Neglecting masses and $t$  compared to
terms going with $s$ or $\qq ^2$, the  Bethe Heitler contribution to the unpolarized
$\gamma p$ cross section is ($M$ is the proton mass) 
\begin{eqnarray}
\label{approx-BH}
\frac{d \sigma_{BH}}{d {Q^\prime}^2 d t d \cos \theta} 
&\approx & 2 \alpha^3 \frac{1}{-t {Q'}^4} \frac{1+\cos ^2 \theta}{1-\cos ^2\theta} 
\left(F_1(t)^2 - \frac{t}{4M_p^2} F_2(t)^2\right) ,
\end{eqnarray}
provided we stay away from the kinematical region where the  product  of lepton propagators goes 
to zero at very small $\theta$ ($F_1(t)$ and $F_2(t)$ are Dirac and Pauli nucleon form factors). The interesting physics program thus imposes a
cut on $\theta$ to stay away from the region where the Bethe Heitler  cross section becomes
extremely large.

\begin{figure}[t]
\begin{center}
    \epsfxsize=0.25\textwidth
     \epsffile{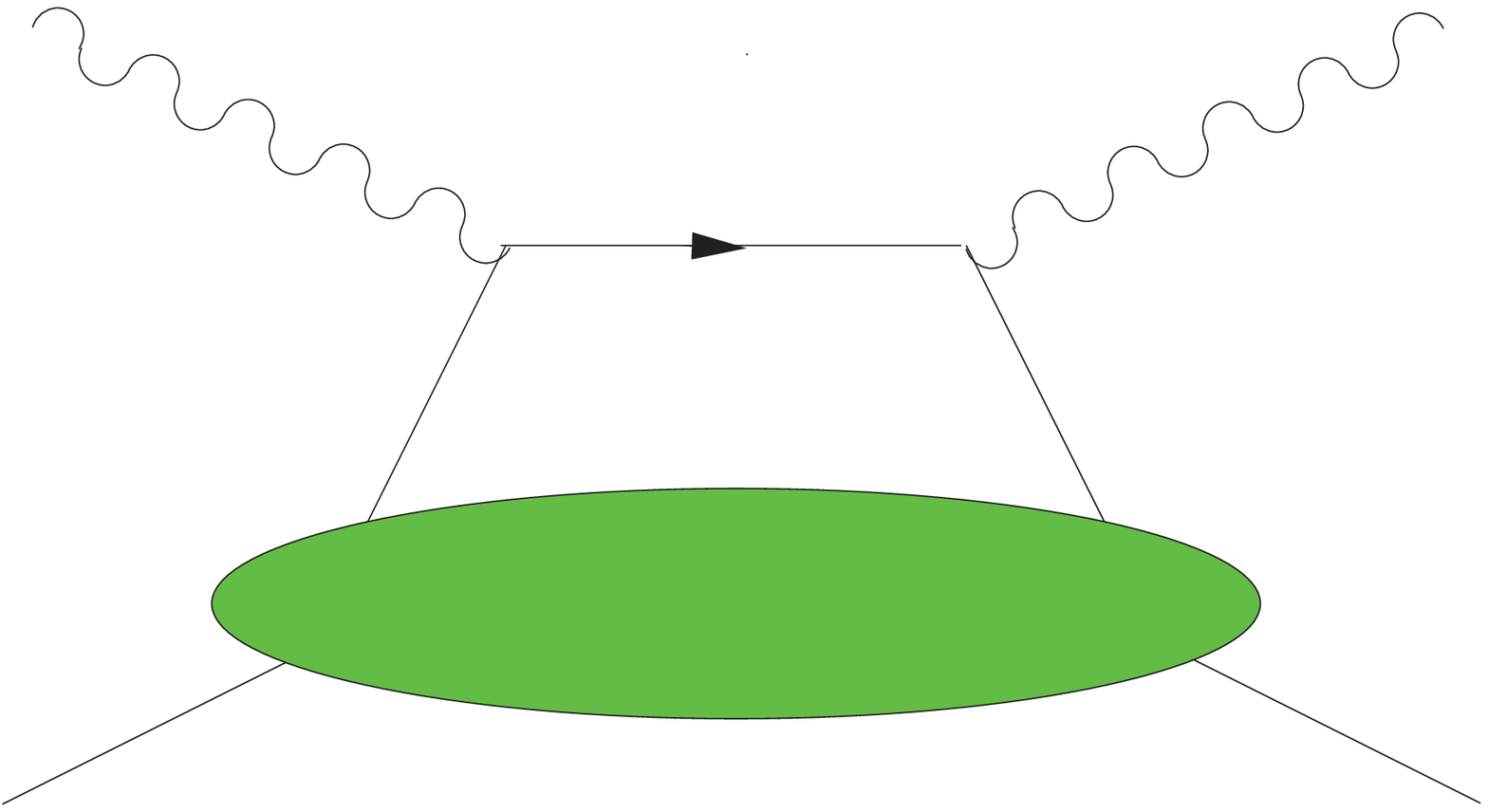}
\hspace{0.05\textwidth}
    \epsfxsize=0.25\textwidth
    \epsffile{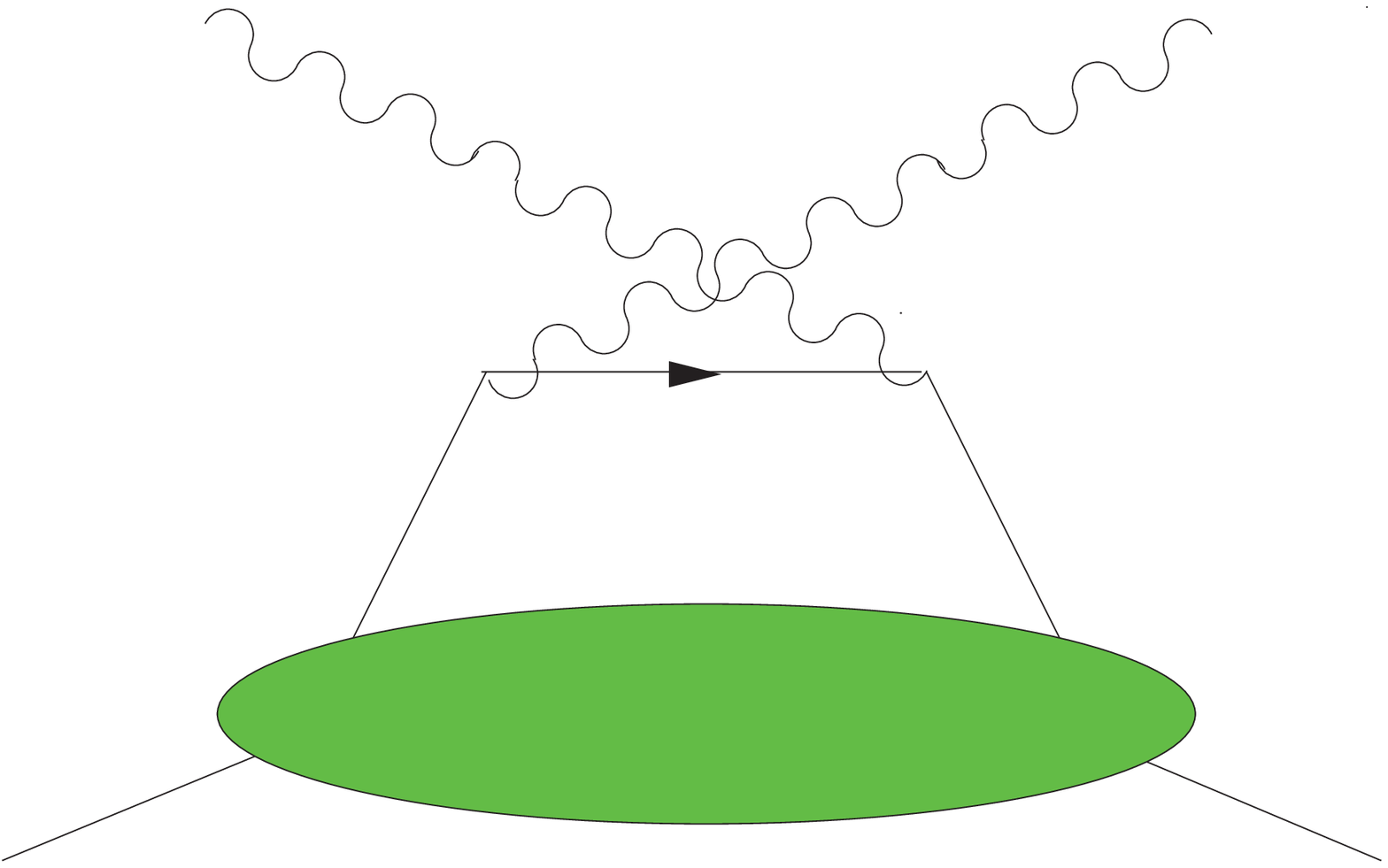}
\caption{Handbag diagrams for the Compton process in the scaling limit.}
\label{haba}
\end{center}
\end{figure}
\begin{figure}[b]
\begin{center}
\epsfxsize=0.4\textwidth
\epsffile{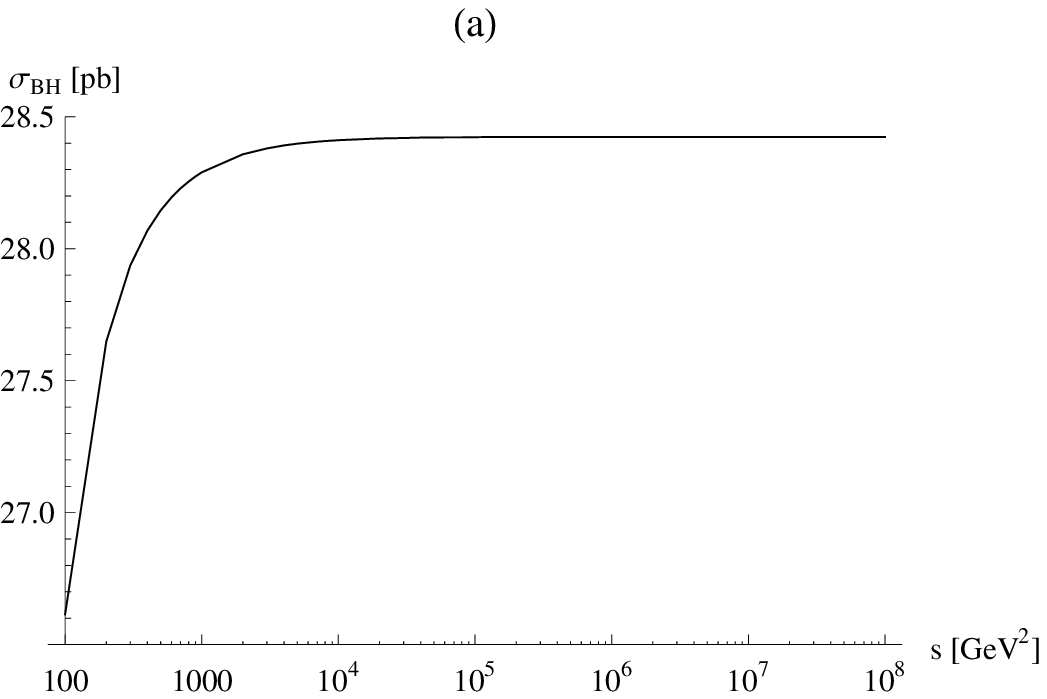}
\hspace{0.05\textwidth}
\epsfxsize=0.39\textwidth
\epsffile{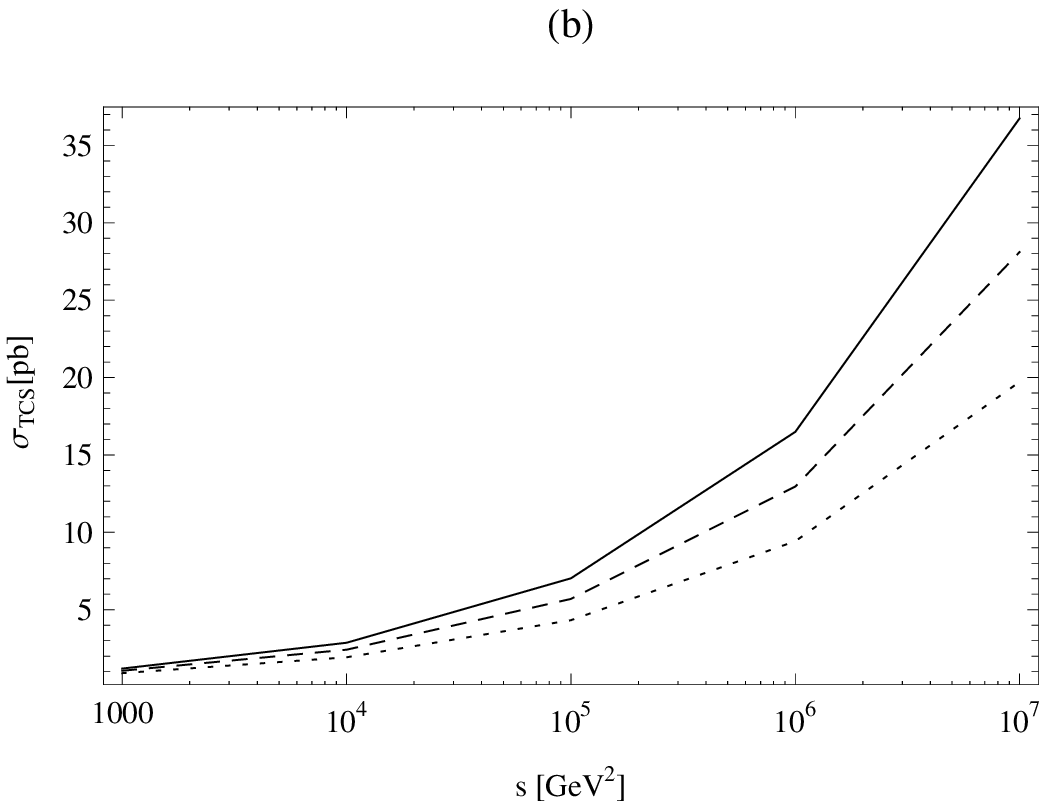}
\caption{ a) The BH cross section integrated over $\theta \in [\pi/4,3
\pi/4]$, $\varphi \in [0, 2\pi]$ , $Q'^2 \in [4.5,5.5]\gev^2$, $|t| \in [0.05,0.25] \gev^2$, as a function of $\gamma p$ c.m. energy squared $s$.
b) $\sigma_{TCS}$ as a function of $\gamma p$ c.m. energy squared $s$, for GPD parametrization based on the 
GRVGJR2008 NLO PDF, for different factorization scales $\mu_F^2 = 4$ (dotted), $5$ (dashed), $6$ (solid) $\gev^2$.
} 
\label{BHs}
\end{center}
\end{figure}
In the region where the final photon virtuality is large, the Compton amplitude is given by the convolution of hard scattering coefficients, calculable in
perturbation theory, and generalized parton distributions, which describe the nonperturbative physics of the process. To leading order
in $\alpha_s$ one then has the dominance of  the quark handbag diagrams of Fig.~\ref{haba}.  
\begin{eqnarray}
\frac{d \sigma_{TCS}}{d {Q^\prime}^2 d \Omega d t} 
\approx \frac{\alpha^3}{8 \pi} \frac{1}{s^2} \frac{1}{{Q'}^2}
\left(\frac{1+\cos^2\theta}{4}\right)
2(1-\eta^2) \left(|\mathcal{H}|^2+|\tilde\mathcal{H}|^2\right) , 
\label{eq:Capprox}
\end{eqnarray}
where $\mathcal{H}$ and $\tilde\mathcal{H}$ are Compton formfactors, defined as in \cite{TCS}, and $\eta$ is the skewedness parameter related to the Bjorken variable $\tau = Q'^2/s$ by $\eta= \tau/(2-\tau)$.
Full BH and TCS cross section as a functions of c.m. energy squared $s$ are shown on Fig. \ref{BHs}. 
Since the amplitudes for the Compton and Bethe-Heitler
processes transform with opposite signs under reversal of the lepton
charge,  
it is possible to project out
the interference term through a clever use of
 the angular distribution of the lepton pair. 
The interference part of the cross-section for $\gamma p\to \ell^+\ell^-\, p$ with 
unpolarized protons and photons is given at leading order by
\begin{eqnarray}
   \label{intres}
\frac{d \sigma_{INT}}{d\qq^2\, dt\, d\cos\theta\, d\varphi}
= {}-
\frac{\alpha^3_{em}}{4\pi s^2}\, \frac{1}{-t}\, \frac{M}{Q'}\,
\frac{1}{\tau \sqrt{1-\tau}}\,
  \cos\varphi \frac{1+\cos^2\theta}{\sin\theta}
     \re\tilde{M}^{--} \; ,
\end{eqnarray}
with ($-t_0 = 4\eta^2 M^2 /(1-\eta^2)$):
\begin{equation}
\label{mmimi}
\tilde{M}^{--} = \frac{2\sqrt{t_0-t}}{M}\, \frac{1-\eta}{1+\eta}\,
\left[ F_1 {\cal H}_1 - \eta (F_1+F_2)\, \tilde{\cal H}_1 -
\frac{t}{4M^2} \, F_2\, {\cal E}_1 \,\right].
\end{equation}

\begin{figure}[t]
\begin{center}
\epsfxsize=0.39\textwidth
\epsffile{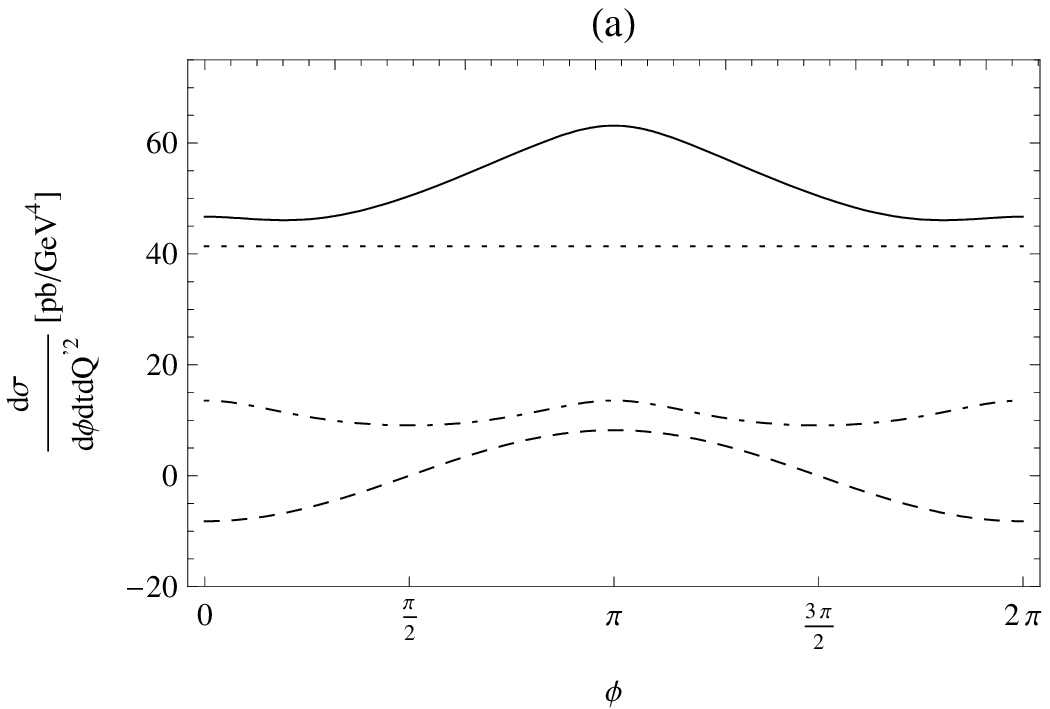}
\hspace{0.05\textwidth}
\epsfxsize=0.39\textwidth
\epsffile{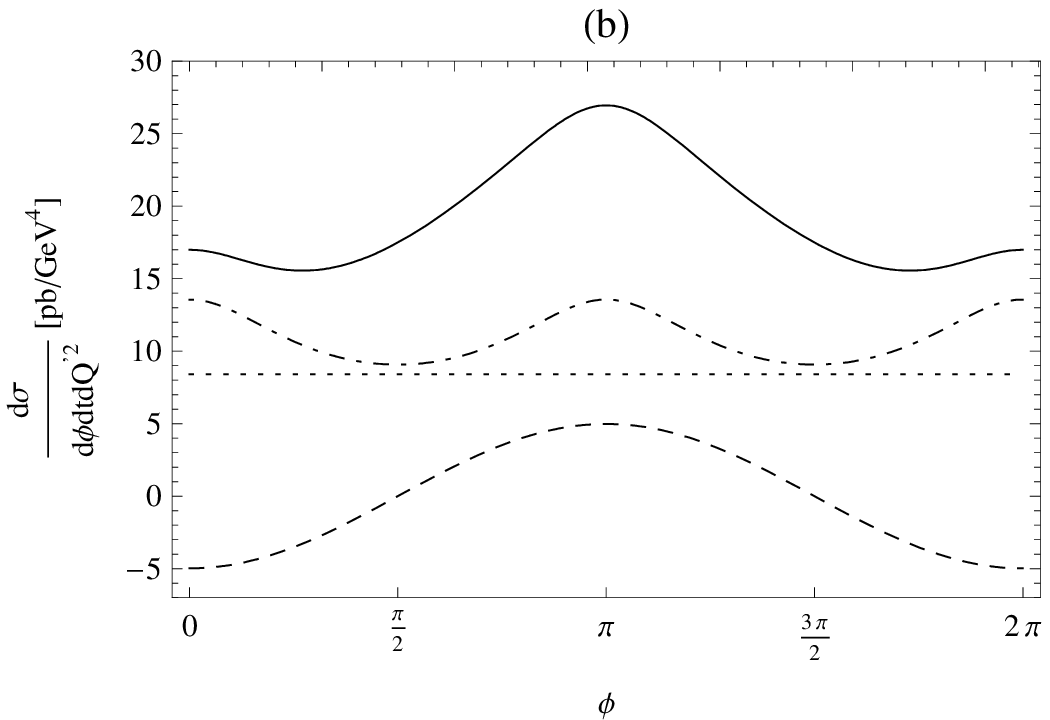}
\vspace{0.05\textwidth}
\epsfxsize=0.60\textwidth
\epsffile{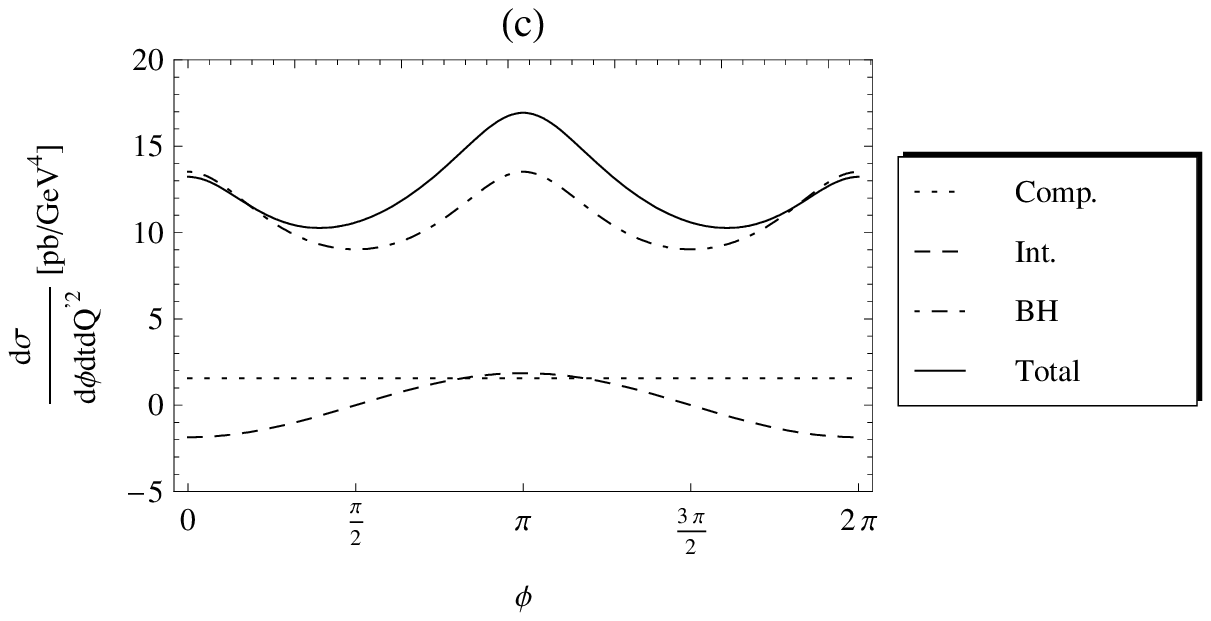}
\caption{
The differential cross sections (solid lines) for $t =-0.2 \gev^2$, ${Q'}^2 =5 \gev^2$ and integrated 
over $\theta = [\pi/4,3\pi/4]$, as a function of $\varphi$, for $s=10^7 \gev^2$ (a), 
$s=10^5 \gev^2$(b), $s=10^3 \gev^2$ (c) with $\mu_F^2 = 5 \gev^2$. We also display  the
Compton (dotted), Bethe-Heitler (dash-dotted) and Interference (dashed) contributions. 
}
\label{Interf}
\end{center}
\end{figure}

In Fig. \ref{Interf} we show the interference contribution to the cross section in comparison to the Bethe Heitler and Compton processes, for various values of c.m. energy squared $s = 10^7 \gev^2,10^5 \gev^2,10^3 \gev^2$. We observe that for large energies the Compton process dominates, whereas for $s=10^5 \gev^2$ all contributions are comparable.
\section{Full cross section}
The cross section for photoproduction in hadron collisions is given by:
\begin{equation}
\sigma_{pp}= 2 \int \frac{dn(k)}{dk} \sigma_{\gamma p}(k)dk
\end{equation}
where $\sigma_{\gamma p} (k)$ is the cross section for the 
$\gamma p \to pl^+l^-$ process and $k$ is the photon energy. 
$\frac{dn(k)}{dk}$ is an equivalent photon flux. 
The relationship between $\gamma p$  energy squared $s$ and k is given by $
s \approx 2\sqrt{s_{pp}}k \nonumber$, 
where $s_{pp}$ is the proton-proton  energy squared ($\sqrt{s_{pp}} = 14 \tev$)

The Bethe - Heitler contribution to $\sigma_{p p}$, integrated over  $\theta = [\pi/4,3\pi/4]$, $\phi = [0,2\pi]$, $t =[-0.05 \gev^2,-0.25 \gev^2]$, ${Q'}^2 =[4.5 \gev^2,5.5 \gev^2]$, and photon energies $k =[20,900]\gev $  gives:
\begin{equation}
\sigma_{pp}^{BH} = 2.9 \pb \;.
\end{equation}  
The Compton contribution (calculated with NLO GRVGJR2008 PDFs, and $\mu_F^2 = 5 \gev^2$) gives:
\begin{equation}
\sigma_{pp}^{TCS} = 1.9 \pb\;. 
\end{equation}
We have choosen the range of photon energies in accordance with expected capabilities to tag photon energies
at the LHC. This amounts to a large rate of order of $10^5$ events/year at the LHC with its nominal 
luminosity ($10^{34}\,$cm$^{-2}$s$^{-1}$). 
\section{Conclusions}
Timelike Compton scattering in ultraperipheral collisions at hadron colliders opens a new way to measure generalized parton distributions. We have found sizeable rates of events at LHC, even for the lower luminosity which can be achieved in the first months of run. Our work has to be supplemented by studies of higher order contributions which  will involve the gluon GPDs.
\vspace{0.5 cm}

\section*{Acknowledgements} This work is partly supported by the ECO-NET program, contract 
18853PJ, the French-Polish scientific agreement Polonium and the Polish Grant N202 249235.


\end{document}